\documentclass[aps,prl,twocolumn,superscriptaddress]{revtex4}

\usepackage{graphicx}
\usepackage{dcolumn}
\usepackage{bm}
\begin{document}

\preprint{APS/123-QED}

\title{Distinct doping dependences of the pseudogap and superconducting gap of La$_{2-x}$Sr$_{x}$CuO$_4$ cuprate superonductors}

\author{M. Hashimoto}
\affiliation{Department of Physics and Department of Complexity Science and Engineering, University of Tokyo, Kashiwa, Chiba 277-8561, Japan}
\author{T. Yoshida}
\affiliation{Department of Physics and Department of Complexity Science and Engineering, University of Tokyo, Kashiwa, Chiba 277-8561, Japan}
\author{K. Tanaka}
\affiliation{Department of Physics and Department of Complexity Science and Engineering, University of Tokyo, Kashiwa, Chiba 277-8561, Japan}
\author{A. Fujimori}
\affiliation{Department of Physics and Department of Complexity Science and Engineering, University of Tokyo, Kashiwa, Chiba 277-8561, Japan}
\author{M. Okusawa}
\affiliation{Department of Physics, Faculty of Education, Gunma University, Maebashi, Gunma 371-8510, Japan}
\author{S. Wakimoto}
\affiliation{QuBS, Japan Atomic Energy Agency, Tokai, Ibaraki, 319-1195, Japan}
\author{K. Yamada}
\affiliation{Institute of Materials Research, Tohoku University, Sendai 980-8577, Japan}
\author{T. Kakeshita}
\affiliation{SRL- ISTEC, Tokyo, 135-0062, Japan}
\author{H. Eisaki}
\affiliation{AIST, 1-1-1 Central 2, Umezono, Tsukuba, Ibaraki, 305-8568, Japan}
\author{S. Uchida}
\affiliation{Department of Physics, University of Tokyo, Tokyo 113-8656, Japan}

\date{\today}

\begin{abstract}
We have performed a temperature-dependent angle-integrated photoemission study of La$_{2-x}$Sr$_{x}$CuO$_4$ covering from lightly-doped to heavily-overdoped regions and oxygen-doped La$_2$CuO$_{4.10}$.
The superconducting gap energy $\Delta_{sc}$ was found to remain small for decreasing hole concentration while the pseudogap energy $\Delta$* and temperature \textit{T}* increase. 
The different behaviors of the superconducting gap and the psudogap can be explained if the superconducting gap opens only on the Fermi arc around the nodal (0,0)-($\pi,\pi$) direction while the pseudogap opens around $\sim$($\pi$, 0).
The results suggest that the pseudogap and the superconducting gap have different microscopic origins.
\end{abstract}


\maketitle
The pseudogap phenomena in the high-\textit{T$_c$} cuprates have been extensively studied since shortly after the discovery of high-\textit{T$_c$} superconductivity because of their peculiarity and possible intimate connection with the unknown mechanism of superconductivity. 
Angle-resolved photoemission spectroscopy (ARPES) studies have revealed the existence of the pseudogap with an energy scale $\Delta^*$, primarily in the antinodal region $\textbf{k}$ $\sim$($\pi$,0) \cite{LSCO_36,LSCO_37,LSCO_2,LSCO_7,LSCO_45,LSCO_44,LSCO_54}. 
Below a characteristic temperature $T^*$, the electrical resistivity \cite{LSCO_25, LSCO_27} and the Cu NMR relaxation rate \cite{LSCO_29} show an anomalous drop  due to the pseudogap opening. 
It has been established that both $\Delta$* and $T$* increase with decreasing hole concentration \cite{LSCO_36,LSCO_37,LSCO_2,LSCO_7,LSCO_11,LSCO_25, LSCO_27,LSCO_29,LSCO_35,LSCO_44,LSCO_45,LSCO_54}. 
Although the origin of the pseudogap is still contraversial, it has been considered that the pseudogap and the superconducting gap may be related and that both the pseudogap and the superconducting gap increase with decreasing hole concentration. 
In the present study, we have found the differnt doping dependences of the superconducting gap and the pseudogap, namely, strong doping dependence of the pseudogap and much weaker doping dependence of the superconducting gap, from detailed photoemission measurements over a wide temperature and doping ranges.

In order to identify the superconducting gap and the pseudogap separately, we note that temperature-dependent photoemission study is a useful method. 
Through angle-integrated photoemission (AIPES) measurements, one can observe the density of states (DOS) and its temperature and doping dependences. 
If a material has an energy gap at the Fermi level, the DOS within a gap energy is expected to decrease below the gap-opening temperature. 
Thus, one may separately observe the superconducting gap and the pseudogap by comparing the DOS at $T$ \textless~$T_c$, $T_c$ \textless~$T$ \textless~$T^*$ and $T$ \textgreater~$T^*$. 
However the detailed studies have been lacking so far on the temperature dependence of gap features in spectroscopic data \cite{LSCO_37,LSCO_1,LSCO_2,LSCO_39}. 
In this work, we have performed a systematic AIPES study of the single-layer cuprates La$_{2-x}$Sr$_{x}$CuO$_4$ (LSCO) and La$_2$CuO$_{4.10}$ (LCO) over a wide temperature and doping range. 
We have succeeded in isolating the superconducting gap $\Delta_{sc}$ from the psudogap $\Delta^*$ by thoroughly analyzing the temperature dependence of the DOS.
In contrast to the general belief that $\Delta_{sc}$ and $\Delta$* are inter-related, the superconducting gap $\Delta_{sc}$ showed no clear doping dependence, while the pseudogap become larger with decreasing hole concentration. 
The observation of the different doping dependence of the $\Delta_{sc}$ and $\Delta$* shall be discussed based on the Fermi arc picture.

\begin{figure*}
\begin{center}
\includegraphics[width=\linewidth]{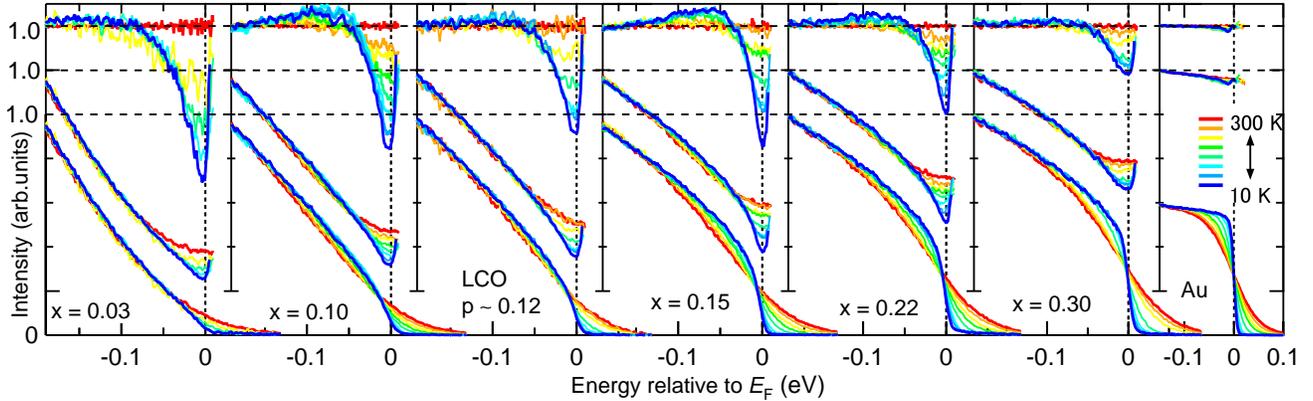}
\caption{(Color online) Temperature-dependent AIPES spectra near \textit{E$_F$} of La$_{2-x}$Sr$_x$CuO$_{4}$ (LSCO) and La$_2$CuO$_{4.10}$ (LCO, $p$ $\sim$ 0.12).
Spectra of gold are shown as a reference.
The lower panel for every doping level shows raw spectra. 
The intensity has been normalized to area under the curve from -0.2 eV to +0.1 eV between different temperatures. 
The middle panel for every sample shows the density of states (DOS) obtained by dividing the spectra by the FD function. 
The upper panels show the ``normalized DOS'' obtained by dividing the DOS by the smoothed DOS at 300 K.}

\label{Int}
\end{center}
\end{figure*}

For LSCO, holes are doped by substituting Sr for La \cite{LSCO_15,LSCO_16,LSCO_17} whereas for LCO, holes are doped by excess oxygen atoms \cite{LSCO_18, LSCO_23}.
The measured samples were LSCO with $x$ = 0.03, 0.10 (\textit{T$_c$} = 25 K), 0.15 (\textit{T$_c$} = 38 K), 0.22 (\textit{T$_c$} = 28 K), 0.30, and LCO with hole concentration \textit{p} $\sim$ 0.12 (\textit{T$_c$} $\sim$ 35 K). 
All the samples were single crystals grown by the traveling solvent floating zone (TSFZ) method.
Temperature-dependent AIPES measurements were performed with a GAMMADATA VUV-5000 light source ($h\nu$ = 21.218 eV) and a SCIENTA SES100 analyzer.
The total energy resolution was set at $\sim$10 meV.
Because of the high stability of the power supply of the analyzer, the accuracy in determining $E_F$ was within 1 meV.
The base pressure of the spectrometer was $\sim2 \times 10^{-10}$ Torr.
All the samples were scraped every 30 minutes in order to avoid surface degradation and to obtain highly reproducible spectra.
We confirmed that the obtained spectra did not show any angular dependence, which ensures that the spectra were angle integrated.
The sample temperature was varied between 10 K and 300 K.
For LCO, the measurements were performed after cooling the samples from room temperature to 10 K at the rate of \textless~0.5 K/min so that the staging of the excess oxygen atoms occurred properly \cite{LSCO_18,LSCO_28}.

The lower panels of Fig.~\ref{Int} show the temperature-dependent AIPES spectra near \textit{E$_F$} for various doping levels.
In the heavily overdoped $x$ = 0.30 sample, the spectral line shapes were similar to those of gold while some asymmetric temperature broadening at \textit{E$_F$} existed.
With decreasing hole concentration, the intensity at \textit{E$_F$} gradually decreased.
For $x$ = 0.15, we obtained the results remarkably \textit{identical} to those reported previously by Sato \textit{et al.} in the entire temperature range \cite{LSCO_1}.
For the lowest doping $x$ = 0.03, which is within the so-called "spin-glass" phase between the superconducting and antiferromagnetic insulating phases, the intensity at \textit{E$_F$} still remained finite, although the overall spectral line shapes were similar to those of an insulator.
The doping dependence of the spectra at the lowest temperature was consistent with the previous photoemission study \cite{LSCO_3}.
For all the doping levels, the intensity at \textit{E$_F$} increased with temperature up to 300 K.
This was most clearly observed in the underdoped samples, and persisted to some extent even at $x$ = 0.30.
In the middle panel of each $x$ in Fig.~\ref{Int}, we show the DOS obtained by dividing the spectra by the Fermi-Dirac (FD) distribution function.
The reliable energy range after the division by the Fermi-Dirac function is within $\sim$3$k_BT$ above $E_F$.
The results clearly show the temperature dependence of the DOS.
In order to demonstrate the relative changes of the DOS with temperature for each doping, we show in the upper panels of Fig.~\ref{Int} the ``normalized DOS'' obtained by further dividing the DOS by the smoothed DOS at 300 K.
The normalized DOS's more clearly reveal the temperature dependence of the pseudogap behavior for each doping level.
All the temperature-dependent normalized DOS revealed pseudogap behavior, even in the $x$ = 0.30 overdoped sample.
The lower the doping level was, the larger the change of the normalized DOS with temperature became.

\begin{figure*}
\begin{center}
\includegraphics[width=1\linewidth]{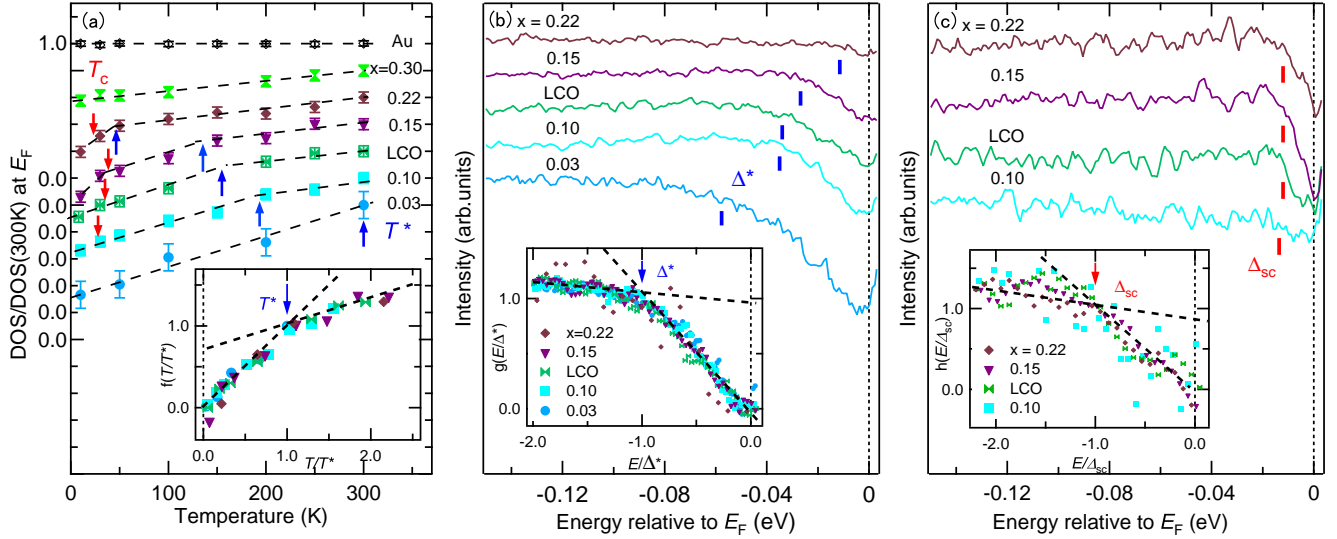}
\caption{(Color online) Temperature and doping dependences of the pseudogap and the superconducting gap in LSCO and LCO.
(a) Normalized DOS at \textit{E}$_F$ [$I_{E_F}(x,T)$] as a function of temperature.
Red and blue arrows denote $T$* and $T_c$, respectively.
Inset shows the scaling plot $I_{E_F}(x,T) = A(x)f(T/T^*)+a(x)$.
Dashed lines are guide for eyes.
(b) Difference normalized DOS [$_{diff}I(x,E)$] between \textit{T} $\textgreater$ \textit{T}* [300 K, 200 K, 200 K, 150 K, 50 K for $x$ = 0.03, 0.10, 0.12 (LCO), 0.15, 0.22, respectively] and $T$* $\textgreater$ \textit{T} $\textgreater$ \textit{T$_c$} [10 K, 30 K, 50 K, 50 K, 30 K for $x$ = 0.03, 0.10, 0.12 (LCO), 0.15, 0.22, respectively] 
Inset shows the scaling plot $I_{diff}(x,E) = B(x)g(E/\Delta^*)+b(x)$ of the difference normalized DOS.
(c) Difference normalized DOS [$I'_{diff}(x,E)$] between $T$* \textgreater \textit{T} \textgreater \textit{T$_c$} [30 K, 50 K, 50 K, 30 K for $x$ = 0.10, 0.12 (LCO), 0.15, 0.22, respectively] and \textit{T} \textless \textit{T$_c$} (10 K).
Inset shows the scaling plot $I'_{diff}(x,E) = C(x)h(E/\Delta_{sc})+c(x)$ of the difference normalized DOS.
}
\label{ScalingAll}
\end{center}
\end{figure*}

In Fig.~\ref{ScalingAll}(a), the normalized DOS at $E_F$, $I_{E_F}(x,T)$, is plotted as a function of temperature for each doping level.
It decreased as the temperature decreased for all the samples.
One can see that the slope becomes large below a certain temperature marked by $T$*.
We attribute this slope change to the opening of the pseudogap below $T$*.
Since all the curves except for $x$ = 0.30 show qualitatively the same behavior, the normalized DOS at $E_F$ $I_{E_F}(x,T)$ can be represented by a scaling formula $I_{E_F}(x,T) = A(x)f(T/T^*)+a(x)$, where $A(x)$ and $I_0(x)$ depend only on the hole concentration $x$, as shown in the inset of Fig.~\ref{ScalingAll}(a).
One can see a kink at \textit{T/T}* = 1, below which $f(T/T^*)$ decreases more rapidly than at $T/T^*$ \textgreater 1.
A possible origin of the slope above $T^*$ is discribed in footnote \cite{LSCO_31}.

\begin{figure}
\begin{center}
\includegraphics[width=\linewidth]{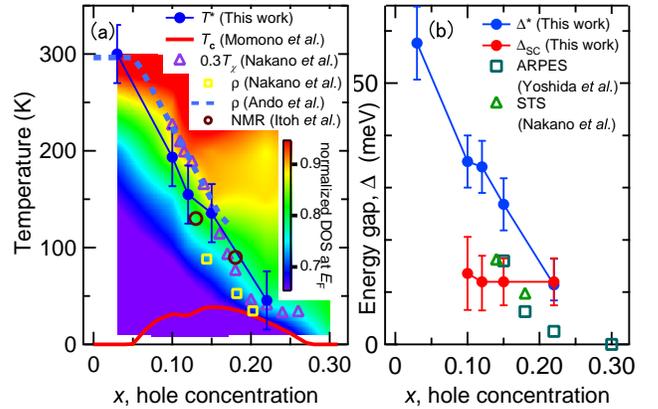}
\caption{(Color) Characteristic temperatures (a) and energies (b) in LSCO and LCO.
(a) \textit{T}* compared with those obtained from the specific heats \cite{LSCO_27}, electrical resistivity \cite{LSCO_25, LSCO_27}, NMR relaxation time \cite{LSCO_29} and magnetic susceptibility \cite {LSCO_24} and \textit{T$_c$} \cite{LSCO_40}.
The color plot shows the normalized DOS at \textit{E$_F$} [$I_{E_F}(x,T)$] and reflects the evolution of the pseudogap in the $x-T$ plane.
(b) \textit{$\Delta$}* and \textit{$\Delta_{sc}$} compared with the leading edge mid point in ARPES spectra at $\sim$$(\pi,0)$ \cite{LSCO_45} and STS \cite{LSCO_27}. 
}
\label{PD}
\end{center}
\end{figure}

In Fig.~\ref{PD}(a), we have plotted the \textit{T}* thus deduced as a function of hole concentration.
The color plot of $I_{E_F}(x,T)$ is also shown in the same panel.
One can see that \textit{T}* increases with decreasing doping level.
The present $T$* agrees well with $T$* previously estimated from the electrical resistivity \cite{LSCO_25, LSCO_27} and NMR \cite{LSCO_29} as plotted in the figure.
It should be noted that the pseudogap behavior in the NMR relaxation rate of LSCO is manifested as a decrease of 1/($T_1T$) and can be explained by the decrease of the DOS because $1/(T_1T)$ is proportional to the DOS at $E_F$.
The unusual drop of the uniform magnetic susceptibility below $\sim$0.3$T_{\chi}$ ($\sim$ $T$*) \cite{LSCO_24} can also be explained by the drop of the DOS at $E_F$ below $T$* observed in the present study.

We derived the magnitude of the pseudogap $\Delta$* by taking the difference $I(x,E)$ between the normalized DOS just above and well below \textit{T}* (but above $T_c$), as shown in Fig.~\ref{ScalingAll}(b).
The pseudogap is reflected in the drop of the difference DOS toward $E_F$ ($E$ = 0).
We defined the kink energy scale of the difference normalized DOS as the pseudogap energy scale $\Delta^*$.
By using a similar scaling formula $I_{diff}(x,E) = B(x)g(E/\Delta^*)+b(x)$, we have obtained the scaled difference DOS $g(E/\Delta^*)$ as shown in the inset of Fig.~\ref{ScalingAll}(b).
Thus obtained $\Delta$* are plotted in Fig.~\ref{PD}(b) as a function of $x$, where $\Delta$* estimated from previous experiments \cite{LSCO_45,LSCO_27} are also plotted.
One can see that \textit{$\Delta$*} increased with decreasing doping level like \textit{T}* and that $\Delta^*/T^*$ is nearly constant $\sim$ (4.3/2)$k_B$, analogous to the $d$-wave gap formula of $2\Delta$ $\sim$ 4.3$k_BT$ \cite{LSCO_30}.

Next, we investigate the signature of the superconducting gap opening in the present spectra.
The normalized DOS have been subtracted between just above \textit{T$_{c}$} (below $T$*) and below \textit{T$_c$} to yield the difference DOS $I(x,E)$ as shown in Fig.~\ref{ScalingAll}(c).
The inset of Fig.~\ref{ScalingAll}(c) shows the scaling analysis of the difference DOS assuming $I'_{diff}(x,E) = C(x)h(E/\Delta_{sc})+c(x)$.
Here $h(E/\Delta_{sc})$ shows the gap feature for \textit{$E$/$\Delta_{sc}$} \textless 1.
We have plotted the obtained \textit{$\Delta_{sc}$} in Fig.~\ref{PD}(b) together with \textit{$\Delta$}* as a function of $x$.
Although the determination of $\Delta_{sc}$ was less accurate than that of $\Delta^*$ because of the  smaller energy scale and the weaker photoemission signals, the different behaviors of $\Delta^*$ and $\Delta_{sc}$ are obvious:
\textit{$\Delta_{sc}$} does not show a clear doping dependence within the present experimental uncertainties.
In particular, \textit{$\Delta_{sc}$} does not show a clear increase with decreasing hole concentration unlike $\Delta^*$.
Although it is difficult to estimate \textit{$\Delta_{sc}$} precisely, one can safely conclude that the doping dependence of the superconducting gap $\Delta_{sc}$ is distinctly  different from that of the pseudogap $\Delta^*$ and remained small compared to $\Delta^*$ in the underdoped region.

The pseudogap opening around ($\pi$,0) has been revealed by many ARPES studies, and from its doping dependence and from its energy and temperature scales, we conclude that the pseudogap observed by AIPES is the same as the pseudogap observed by ARPES .
Using the pseudogap picture observed by ARPES experiments and our observation that there was another gap behavior which showed different doping dependence from the pseudogap and could be attributed to the superconducting gap, we discuss the possible scenario of gap formation using the Fermi arc picture.
The different behaviors of $\Delta_{sc}$ and $\Delta^*$ may be explained if the effective superconducting gap in the underdoped samples opens only on the Fermi arc around the nodal (0, 0)-($\pi$, $\pi$) direction, where low energy excitations are possible, while the pseudogap opens primarily around $\sim$($\pi$, 0) and apart from the Fermi arc on the much higher energy scale.
The idea of the above picture has been suggested by Oda \textit{et al.} \cite{LSCO_33}. 
If this is the case, the length of the Fermi arc will exhibit a stronger temperature dependence at low temperatures in addition to the recent observation that the Fermi arc length is proportional to \textit{T}/\textit{T}* \cite{LSCO_35}. 
This has to be tested in future temperature-dependent ARPES measurements on LSCO.
Very recently, Tanaka $et$ $al$. \cite{LSCO_38} has reported for deeply underdoped Bi2212 that there are two different energy gaps around the node and around $\sim$($\pi$, 0), respectively, and that they show apparently opposite doping dependences.
In Raman experiment, Tacon $et$ $al.$ \cite{LSCO_42} have also revealed that there are two different energy scales in the nodal $B$$_{2g}$ and antinodal $B$$_{1g}$ components for electronic excitations in underdoped cuprates.
The different doping dependences of $\Delta_{sc}$ and $\Delta^*$ suggest that the superconducting gap on the Fermi arc and the pseudogap around $\sim$$(\pi,0)$ may have different origins and are competing with each other.
The pseudogap may be originated from antiferromagnetic fluctuations \cite{LSCO_50} or charge ordering \cite{LSCO_46} as suggested for Ca$_{2-x}$Na$_x$CuO$_2$Cl$_2$. 

In conclusion, we have performed a systematic temperature-dependent AIPES study of LSCO and LCO in a wide doping range. 
We have found that the doping dependence of \textit{$\Delta_{sc}$} is much weaker and remains considerably smaller than that of $\Delta^*$ in the underdoped samples. 
The different doping dependences of \textit{$\Delta_{sc}$} and $\Delta^*$ are consistent with the Fermi arc picture and suggest that the stage of the superconductivity in the underdoped region is primarily on the Fermi arc around the node. 
Also, the contrasting behaviors of \textit{$\Delta_{sc}$} and $\Delta^*$ imply that the pseudogap around the antinode depress superconductivity and has a different microscopic origin from the superconducting gap, possibly antiferromagnetic fluctuations \cite{LSCO_50} or charge ordering \cite{LSCO_46}. 
The relationship between the present observation and the Fermi arc picture should be directly confirmed by temperature-dependent ARPES experiments in future.
 
This work was supported by a Grant-in-Aid for Scientific Research in Priority Area ``Invention of Anomalous Quantum Materials'' from MEXT, Japan.
Informative discussion with Z.-X. Shen is gratefully acknowledged.

\end{document}